# Four-electrodes DBD plasma jet device with additional floating electrode

Fellype do Nascimento [1 §], Munemasa Machida [2], Konstantin G. Kostov [3], and Stanislav Moshkalev [1]

[1] Center for Semiconducting Components and Nanotechnologies – State university of Campinas, Campinas, Brazil

[2] Institute of Physics “Gleb Wataghin” – State University of Campinas, Campinas, Brazil

[3] Faculty of Engineering – Sao Paulo State University, Guaratingueta, Sao Paulo, Brazil

## Abstract

A Dielectric Barrier Discharge (DBD) plasma jet in a four electrodes configuration was investigated in order to improve the discharge parameters, such as, plasma power and rotational and vibrational temperatures of molecular species in the plasma plume. The improvement attempts were made by introducing an auxiliary floating electrode in a form of a metallic pin inside the DBD device. That piece was placed near the bottom of the main device, centered in relation to the four powered electrodes, which were covered with a dielectric material. By using metallic pins with different lengths, it was observed that there were considerable variations of the plasma parameters as a function of the pin length. Two carrier gases were tested: argon and helium. With helium as the working gas, it was found that there is an optimal pin length that maximizes the plasma power and its vibrational temperature. In addition, it was verified that for the pin of optimum length the relative intensity of light emissions from OH and NO species achieved higher values than in other conditions studied.

## 1 Introduction

Atmospheric Pressure Plasma Jets (APPJs) are kind of gas discharges that can be produced in open environments, at room temperature and pressure. Plasma plumes of a few centimeters in length can be obtained using a wide variety of APPJ devices that can be built in different way. In most cases the discharges are excited in a noble gas that flows through a dielectric tube and the resulting plasma is injected into the surrounding ambient [1, 2, 3, 4, 5, 6, 7]. Even though the gas temperature of the plume can be kept close to room temperature, some energetic plasma species (electrons and metastables) interact with air molecules producing different reactive species, mainly reactive oxygen (RO) and nitrogen (RN) species, neutrals in fundamental and excited states and energetic photons [7, 8, 9, 10, 11].

§ Corresponding author: fellype@gmail.com

Along the last decades the use of Dielectric Barrier Discharge (DBD) plasmas has grown a lot and many applications have been developed [12, 13, 14]. Currently, there is a great variety of DBD devices that are employed to produce APPJs most of them being in cylindrical geometry. The most common configurations are using ring electrodes around a dielectric pipe or a combination of pin-ring electrodes where high voltage (HV) is applied to the pin electrode while the ring one is grounded, being that in the last case the pin electrode can be covered by a dielectric material (encapsulated) or bare (exposed) [5, 14]. Both have their advantages and disadvantages comparing with each other. For example, the exposed electrode case requires a much lower voltage to produce a plasma jet than the encapsulated one. On the other hand, due to electrical insulation provided by the dielectric material, encapsulated electrodes are more suitable for *in vivo* applications.

Thus, for specific applications where encapsulated DBD plasma jets are required, it would be interesting to find device configurations and/or operation conditions in which the plasma jet achieve appropriate parameters. For instance, some materials require more plasma power in order to get their surface modified or activated [15, 16]. It is also known that the gas vibrational temperature ($T_{vib}$) of an APPJ plays an important role in the process of surface activation, so that plasma jets with higher $T_{vib}$ values would be more efficient [17, 18, 19]. Lower gas temperatures may be desired in order to avoid possible overheating and thermal damage on the treated material. Furthermore, it has been reported that RO and RN species generated in APPJs play an important role when these kind of plasmas are applied to cancerous tissues, being that, in general, such species react with the free radicals present in those tissues, destroying its diseased cells, but preserving the healthy ones [20, 21, 22, 23, 24, 25, 26, 27, 28].

Previous studies compared the differences in properties of APPJs produced using He as the working gas with and without a target (sample or substrate) in front of the plasma jet [29, 30, 31, 32]. In all these studies there have been considerable increases in density of He atoms in metastable states ($He^m$) as well as in the density of reactive species in the plasma, in the condition where the target was a conductive material. Zaplotnik *et al*, in special, performed experiments using targets with different conductivity values, and it was verified that the higher the conductivity, the higher the $He^m$ density in the plasma jet [31].

In a previous work of our group [33], it was demonstrated that the use of a DBD device in a four-electrodes configuration improves the energy efficiency when compared to cases where less electrodes are employed.

In this work, we further explore the four-electrodes configuration by inserting a metallic pin inside the DBD device that acts as an auxiliary floating electrode (AFE). The AFE is positioned at the bottom of the device, near to the gas outlet, and centered in relation to the four powered electrodes. Thus we investigated the behavior of plasma power, vibrational and rotational temperatures, as well as the production of RO and RN species in the plasma jet – mainly hydroxyl (OH) and nitric oxide (NO) – all them as a function of the relative length ($\Lambda$) of the AFE (length of the pin in relation to the internal length of the DBD device). In the study it was found that there is a pin length value $\Lambda_0$ that seems optimizing the plasma power, $T_{vib}$ and production of reactive species, when helium (He) was used as the working gas. When argon (Ar) was used, there is no clear evidence that there is a value of $\Lambda$ that optimizes the plasma parameters. However, coincidentally the lowest rotational temperature ($T_{rot}$) value obtained with Ar occurred when the $\Lambda$ value was near to the $\Lambda_0$ value obtained for He.

# 2 Experimental setup and procedure

The effects of insertion of a floating electrode into a four-electrodes DBD plasma device were investigated using electrical and optical measurements. The schematic of the DBD device is shown in Fig. 1. It consists of a dielectric enclosure made of nylon, in which are placed four 2.5-mm-diam. copper rods encapsulated in closed-end glass tubes. The main device dimensions are: dielectric enclosure inner and outer diameter 24 and 39 mm, respectively, internal length ($L_i$) 50 mm and outer diameter of the glass tubes is 4.3 mm. The four electrodes are arranged in a square shape with a side equal to 7.0 mm. The main dielectric enclosure terminates with an exchangeable nozzle, which in turn has a plastic support on which metal pins can be fastened and changed in order to act as an auxiliary floating electrode (AFE). The AFE material is a low-carbon steel.

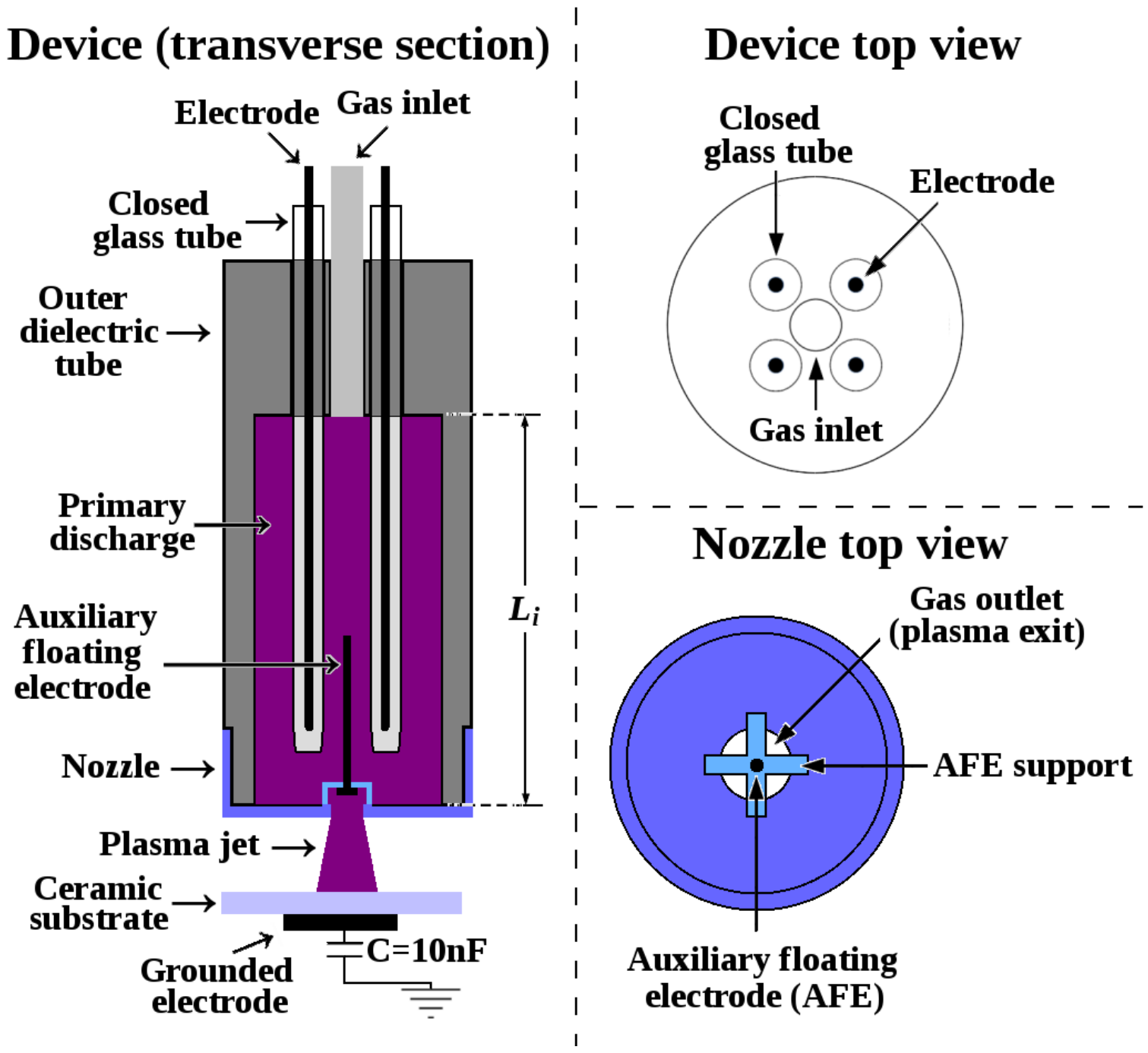

**Figure 1:** Scheme of the DBD device showing the transverse section and top view of the device, and the top view of the nozzle. $L_i$ = 50 mm. The elements are out of scale.

In order to produce a plasma jet, a continuous gas flow is injected inside the outer dielectric tube and the same alternate high voltage is applied to all four electrodes. When the plasma is ignited the primary discharge is mainly concentrated around the AFE, which does not interfere with the exit of the plasma into the environment. The jet was directed to a grounded conductor disc with radius 5 cm that is covered by a 12-cm-diam ceramic substrate. The distance between the ceramic substrate and exit nozzle was kept 12 mm.

In the scheme shown in Fig. 1, the only variable parameter was the length $l_p$ of the metallic pin (the AFE). For this purpose, we used a set of eight pins with lengths ranging from 0.0 mm (no pin) to 37.4 mm, with proportional lengths Λ ($\Lambda = l_p/L_i$) ranging from 0.0 to ~0.77 (0 to ~77%). For each pin used, we recorded the corresponding emission spectrum and measured the high voltage signal applied to the four electrodes as well as the voltage across a serial capacitor $C$ connected to the grounded substrate. The power source was a Minipuls4 generator (Dresden, Germany), the high-voltage signal was kept constant in both amplitude (20kV peak to peak) and frequency (10kHz).

To obtain the optical emission spectra, a multi-channel spectrometer from Avantes (model AvaSpec-ULS2048X64T), with full width at half maximum (FWHM) of (0.678 ± 0.022) nm was used. The light emitted by the plasma jet were collected using a cosine lens, with 10 mm in diameter and placed 6.0 cm from the center of the plasma column, thus giving us a collection area with dimensions of approximately 10.0 mm x 5.0 mm, integrated along a line of sight through the plasma column. After that, the light was transported to the spectrometer through an optical fiber. The spectra were acquired in a continuous mode using an integration time of 25 ms in all cases. A relative calibration of intensity of the spectrometer was performed to check if there was a significant variation in the sensitivity within the narrow wavelength range (from 360 to 385 nm) used to calculate $T_{rot}$ and $T_{vib}$, and it only a small variation (less than 1%) was found.

The plasma power calculations were performed using a well-known method that relates the plasma power to the area of the Lissajous figure formed between voltage ($V$) applied to the device and the charge ($Q$) transferred to the capacitor $C$, with the mathematical expression for the power ($P$) given by [34, 35]:

$$P = f \oint V \, dQ \qquad (1)$$

where $f$ is the frequency of the high-voltage signal.

The plasma temperatures $T_{rot}$ and $T_{vib}$ were obtained using a well-accepted method that compare experimental data of the second positive system emissions from molecular nitrogen ($N_2$ I, C $^3\Pi_u$,v' → B $^3\Pi_g$,v'') with simulated ones [2, 36, 37]. The simulated spectra were generated for various data-sets of $T_{rot}$ and $T_{vib}$ using data from SpecAir software in order to find the one that best fit the experimental spectra in each case, with a coefficient of determination ($R^2$) closer to unity [38]. Then the uncertainty $\sigma T$ ($T = T_{rot}$ or $T_{vib}$) for each temperature was defined taking into account the step $\Delta T$ used to generate the data-sets and the $R^2$ value as:

$$\sigma T=\sqrt{[\Delta T/2]^2 + [(1-R^2)\cdot T]^2} \qquad (2)$$

This was made in the wavelength range from 360 to 385nm because in this range, depending on the working gas used to produce the plasma and vibrational temperature reached, up to five emission peaks from ($N_2$ I, C $^3\Pi_u$,v' → B $^3\Pi_g$,v'') can be observed, and this favors more accurate temperature measurements.

The working gases used to produce the plasma jets were Ar and He, both with purity of 99.98%, and a flow rate of 1.5 L/min in all cases.

# 3 Results and discussion

In this section we present the results obtained from the measurements of plasma power, rotational and vibrational temperatures and the relative intensities of light emissions from reactive species compared with the emissions from excited $N_2$ molecules, as well as an overview of the emission spectra and some details of them.

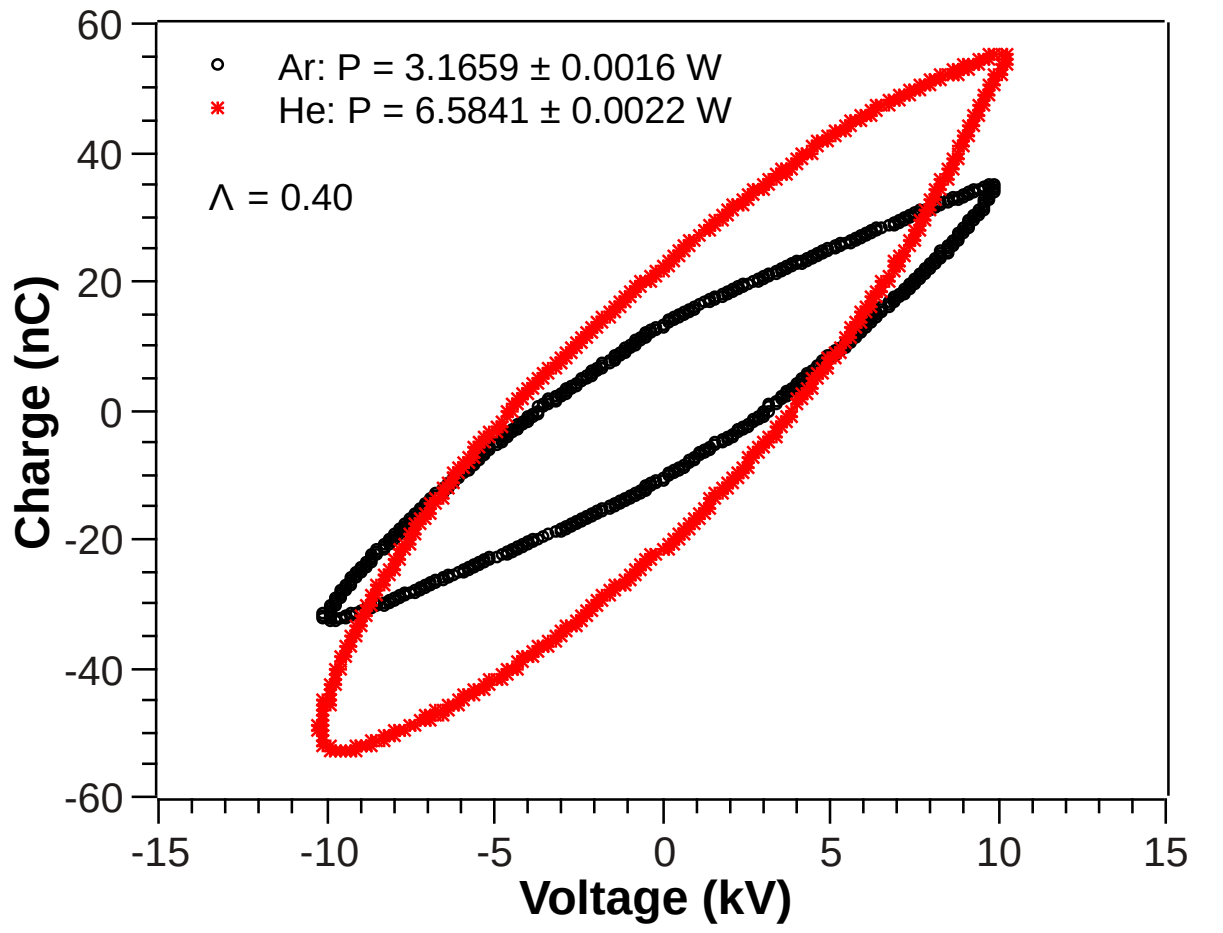

**Figure 2:** Examples of Q-V Lissajous curves used for power calculations for argon and helium. Λ ≈ 0.40 in both cases.

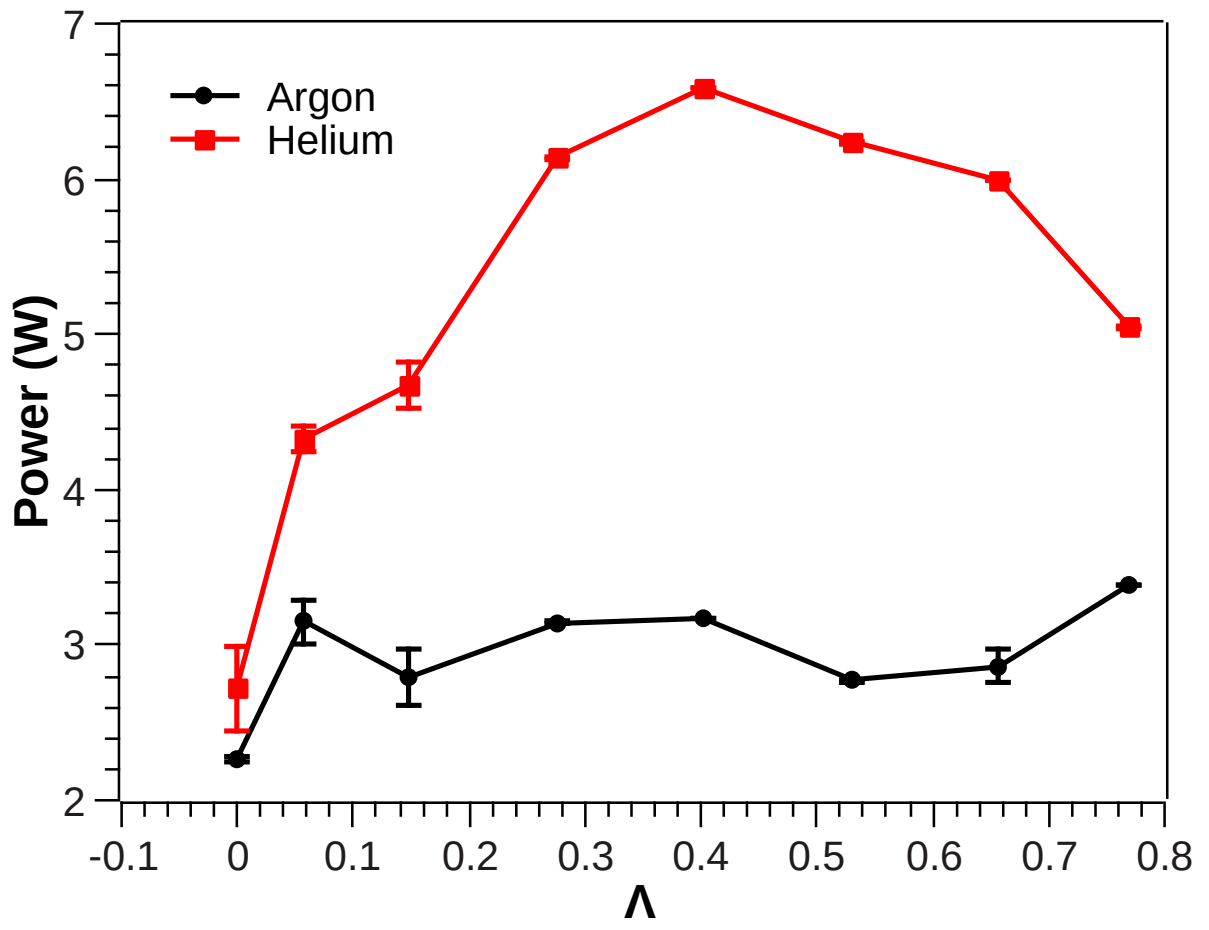

**Figure 3:** Plasma power as a function of Λ for argon (circles) and helium (squares) as the working gases.

## 3.1 Power measurements

Figure 2 shows examples of Q-V Lissajous figures used for plasma power calculation for Ar and He as working gases. In those cases, the proportional length Λ of the pin was ~0.40. The behavior of plasma power as a function of Λ for Ar and He is shown in Fig. 3. As can be observed in Fig. 3, when He is used the values of plasma power show noticeable variations with Λ, presenting a maximum at Λ ≈ 0.40. This is the best value of Λ to maximize the power extraction from the DBD device in the He case. In striking contrast, when Ar was used as a working gas, despite the initial rise clearly observed at Λ ≈ 0.06, further increase of Λ did not result in a stable increase of plasma power.

In Fig. 3 it can be also seen that the use of an auxiliary floating electrode always leads to higher plasma power than with no pin, even at the smallest length. One of possible explanations for this finding is that when the metallic AFE is subjected to the plasma potential, it emits a considerable amount of electrons which increases the probability of formation of streamers in the plasma jet, resulting in discharges with greater power.

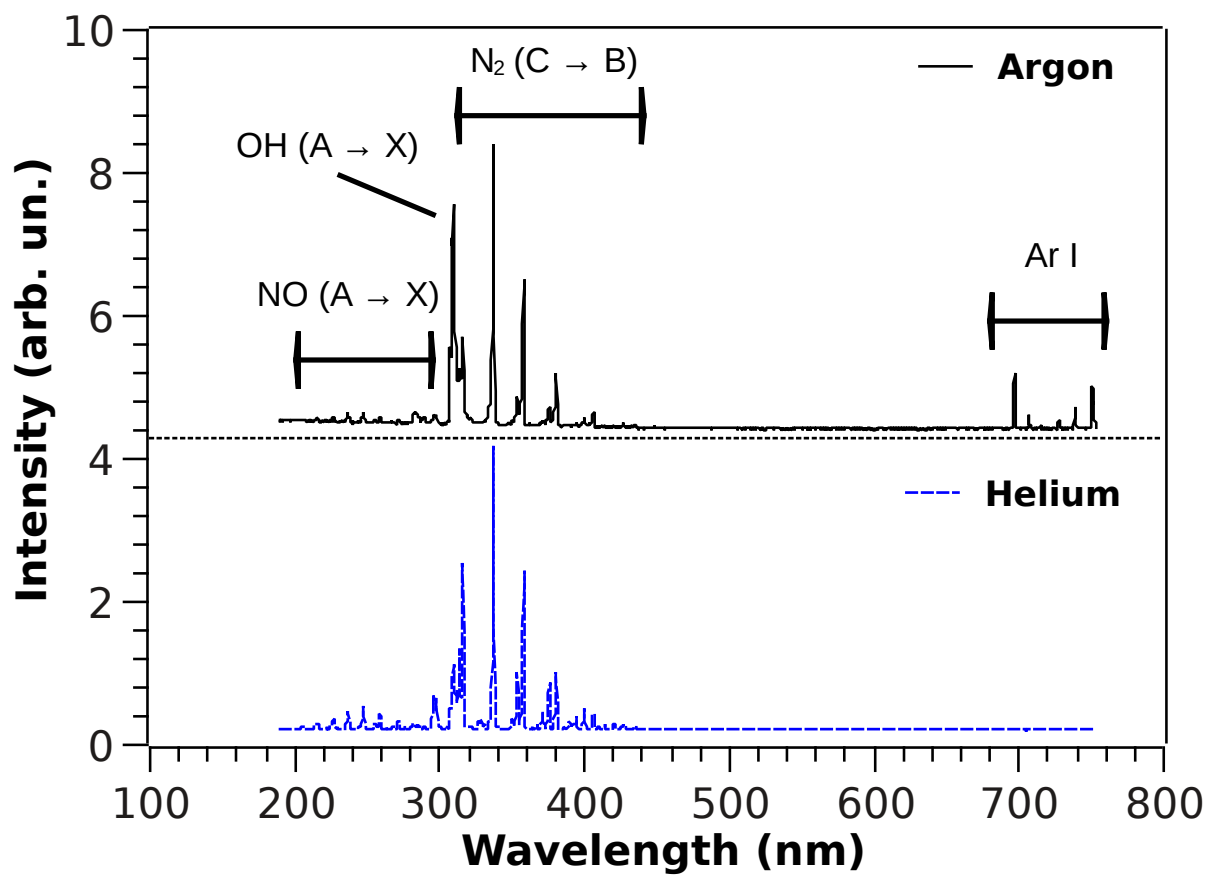

**Figure 4:** Emission spectra of plasma jets formed using argon (above) and helium (below) as working gases. The main emitting species identified are indicated.

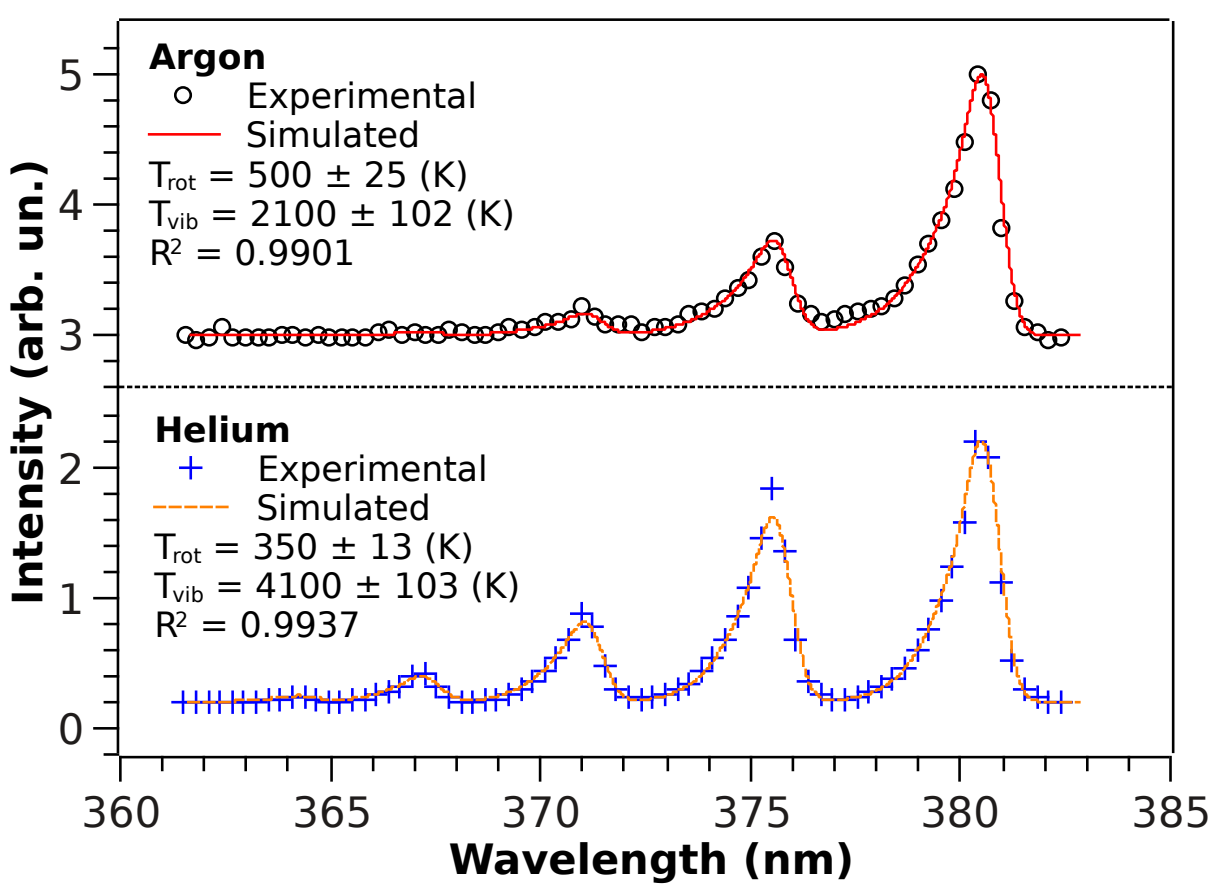

**Figure 5:** $N_2$ I (C → B) emissions in the wavelength range from 360 to 385 nm, and their corresponding simulated spectra that best fit the experimental data for argon (above) and helium (below) as the working gases. Λ = 0.40 in both cases.

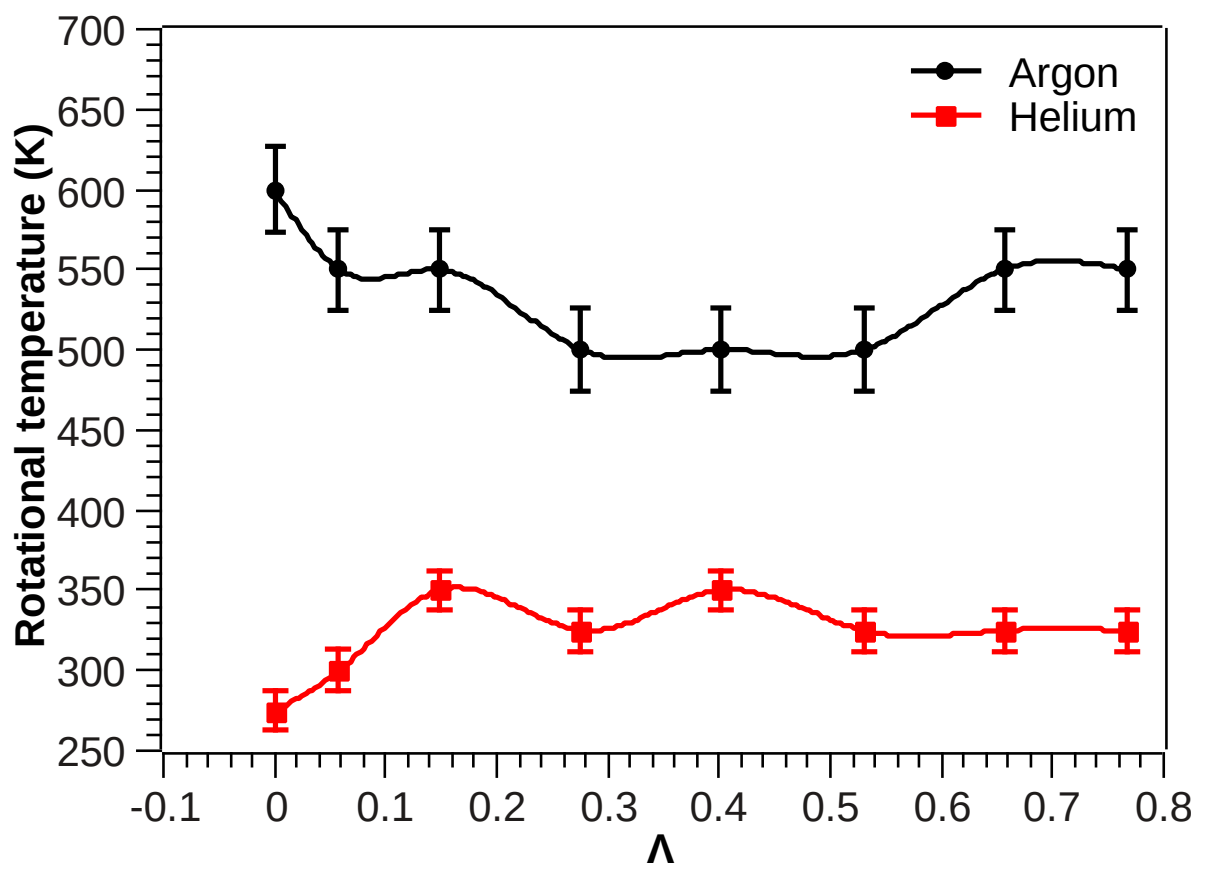

**Figure 6:** Behavior of rotational temperatures as a function of Λ using argon or helium as working gases.

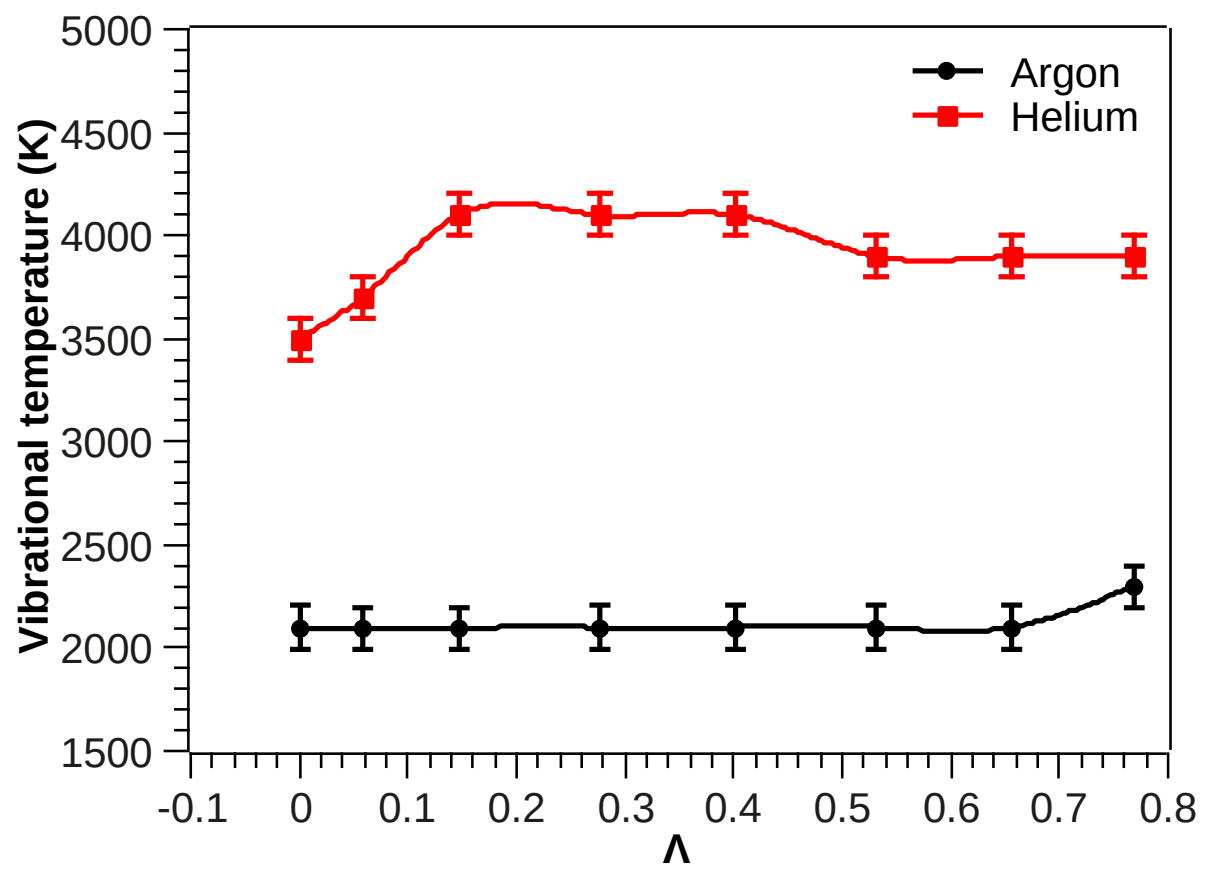

**Figure 7:** Behavior of vibrational temperatures as a function of Λ using argon or helium as working gases.

## 3.2 Temperature measurements

Typical emission spectra of DBD plasma jet, in the wavelength range between 200 and 750nm, using Ar and He are shown in Fig. 4. Examples of $N_2$ (C → B) emissions in the wavelength range from 360 to 385nm, with their corresponding simulated spectra that best fit the experimental data are shown in Fig. 5 for both Ar and He. Figures 6 and 7 show the behavior of $T_{rot}$ and $T_{vib}$, as functions of Λ respectively.

In the emission spectra of both Ar and He plasma jets the main emitting species are NO (A → X), OH (A → X) and $N_2$ (C → B). Characteristic Ar emissions appears only for Ar plasma jet. On the other hand $N_2^+$ (B → X) emission is observed only when He is used.

The $\nu' \rightarrow \nu''$ transitions from $N_2$ I (C → B), ranged from the most intense peak to the lowest one, are: $0 \rightarrow 2$, $1 \rightarrow 3$, $2 \rightarrow 4$, $3 \rightarrow 5$, and $4 \rightarrow 6$. In He discharges all five peaks are observable and can be used to calculate $T_{vib}$, but when Ar is used in general the number of peaks that can be observed is only 3. This is likely due to the lower $T_{vib}$ values achieved in Ar discharges.

As can be seen in Fig. 6, the $T_{rot}$ values after the inclusion of the AFE (Λ > 0) are almost constant in relation to variation of Λ for both He and Ar, but for the former a minimum level for Λ between ~0.3-0.5 is clearly observed. Fig. 7 shows that the Λ value does not affect the behavior of $T_{vib}$ in Ar discharges. On the other hand, when He is used, the $T_{vib}$ values change considerably depending on the Λ value. The $T_{vib}$ curve for He in Fig. 7 shows values ranging from 3900 K to 4100 K for Λ ranging from ~0.14 to 0.77, which indicates that there are Λ values that are better to obtain higher $T_{vib}$ values.

## 3.3 Increment in the relative number of reactive species

The addition of the auxiliary floating electrode in the device allowed us to study the variation in the relative number of reactive species (essentially OH and NO) in the plasma jet as a function of Λ. As the intensity of spectroscopic emissions ($I_{species}$) from atoms and/or molecules in a plasma is proportional to the density of such particles ($n_{species}$), that is $I_{species}\ \alpha\ n_{species}$, we can say that the larger are the ratios $I_{\mathrm{OH}}/I_{N_2}$ and $I_{\mathrm{NO}}/I_{N_2}$, the greater is the number of the reactive species OH and NO in excited states in the plasma jet in comparison to the number of $N_2$ molecules. The band emissions used to measure the intensity ratios were those resulting from $\nu'=0 \rightarrow \nu''=0$ transitions (0–0 transitions), whose wavelengths are 337.13 nm, 308.90 nm and 226.28 nm for $N_2$, OH and NO, respectively. The 0–0 transitions were chosen for intensity ratio calculations because they are usually those with the greatest transition probability and also to maintain a consistent criterion through all comparisons. Figures 8 and 9 show the behavior of intensity ratios as a function of Λ for cases where Ar and He were used as working gases, respectively.

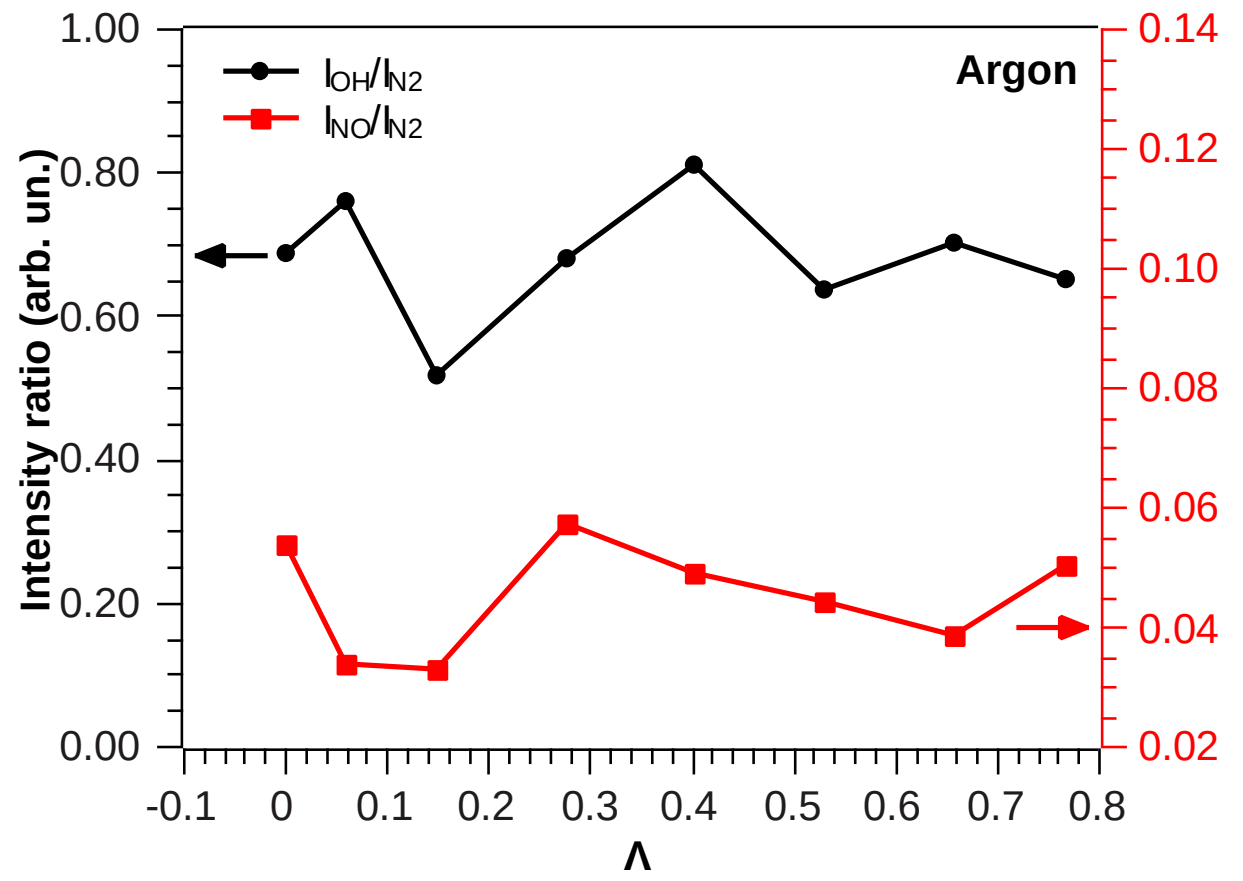

**Figure 8:** Intensity ratios between reactive species OH and NO in relation to $N_2$ molecules as a function of Λ in argon discharges.

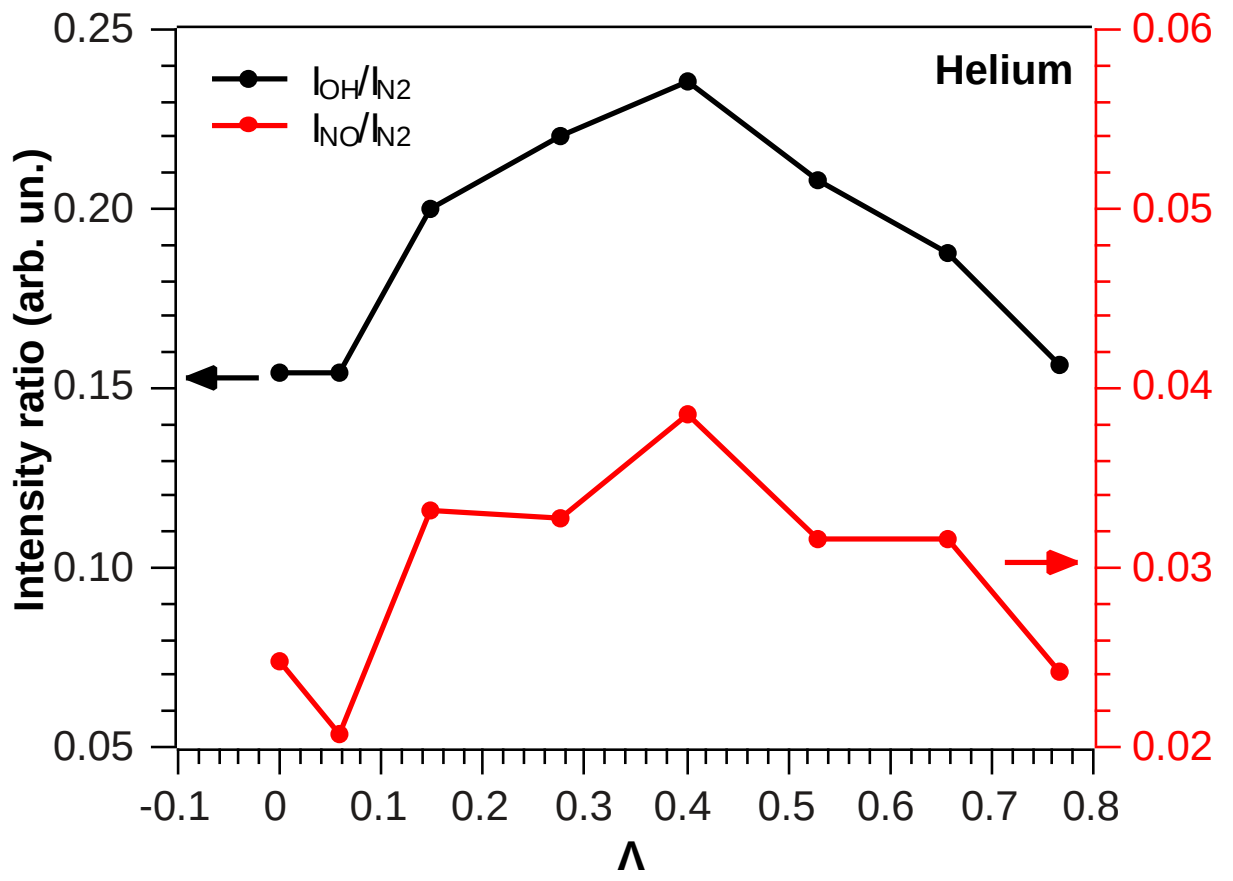

**Figure 9:** Intensity ratios between reactive species OH and NO in relation to $N_2$ molecules as a function of Λ in argon discharges.

From Fig. 8, can be seen that the production of excited reactive species in the Ar plasma jet does not look to have a strong correlation with the Λ value. However, when He is used to produce the plasma jets there is an increment in the production of both OH and NO when Λ is increased from 0 to about 0.40, where the production of the reactive species reaches its maximum value. Then, as Λ continues to increase, both OH and NO species are reduced. This kind of dependence on Λ in OH and NO production only when He is used correlates well with the behavior of plasma power as a function of Λ (see Fig. 3). Once all the other conditions to produce the plasma have been kept constant, the changes in the relative number of both OH and NO molecules are likely to be related to ‘an exchange’ of the energy from $N_2$ molecules to those as Λ is altered.

Since the plasma power and production of reactive species present almost the same dependence on the Λ value only for He gas, we can infer that there is some coupling between the electric field formed around the floating electrode that is only allowed to happen in He discharges.

In principle, metals under a plasma bombardment have a large probability to release both electrons and/or metallic ions to the plasma jet, increasing the contribution of the electron impact to the excitation of reactive species and also generating a higher electric current and increasing the plasma power. Then, it would be expected to have more electrons released from the auxiliary floating electrode with Λ being increased, which, in turn, would result in higher values of intensity ratios for higher Λ values. However, this was not observed in practice, as can be seen in Fig. 8, for operation using Ar, and in Fig. 9, for operation with He. Instead we see that when using He, there is an ideal Λ value that provides a higher production of reactive species in plasma. In addition, the increase in OH production in the presence of the AFE shown in figure 9 is in agreement with what has already been reported previously in the literature [29, 30, 31, 32].

# 4 Conclusions

From the results of plasma power, vibrational temperatures and intensity ratios between emissions of reactive species and molecular nitrogen, it can be stated that the addition of the AFE in the four-electrodes DBD device leads to enhancement of such plasma parameters when He is used as the working gas. In addition, it was possible to verify that there is an appropriated length of the AFE that seems to optimize the plasma parameters, and it was found to be ~40% of the internal depth of the DBD device ($\Lambda_0 \approx 0.40$). On the other hand, when Ar is used as the working gas the AFE does not seem to have any practical effects on most of the plasma parameters concerned in this work, but, coincidentally, the lowest rotational temperature recorded in operations with Ar gas was found for $\Lambda \approx \Lambda_0$.

Further works will include the influence of the AFE material in plasma properties and its proportional diameter. Furthermore, a study using DBD devices with diameter and length other than the used in this work can be used in order to verify if the role of the proportional AFE length in DBD plasma parameters is almost the same under all operation conditions.

# Acknowledgments

Part of this work was supported by CNPq.